\documentclass[a4paper,12pt]{article}

\usepackage{bbm}
\usepackage[utf8]{inputenc}
\usepackage{placeins}
\usepackage{amssymb,amsmath,amsfonts}
\usepackage[colorlinks,linkcolor=purple,citecolor=teal]{hyperref}
\usepackage{dsfont}
\usepackage{color}
\usepackage{graphicx}
\usepackage{hyphenat}
\usepackage{wrapfig}
\usepackage{empheq}
\usepackage{textcomp}
\usepackage[caption=false]{subfig}
\usepackage{rotating}
\usepackage{wrapfig}
\usepackage{pdfpages}
\usepackage{verbatim}
\usepackage{setspace}
\usepackage{array}
\usepackage{caption}
\usepackage[pdf]{pstricks}
\usepackage{upgreek}
\usepackage{cmll}
\usepackage{latexsym}
\usepackage{braket}
\usepackage{cite}
\usepackage{epsfig}
\usepackage[left=2.4cm,top=3.3cm,right=2.4cm,bottom=3.3cm,bindingoffset=0cm]{geometry}
\usepackage[titletoc,page]{appendix}
\usepackage{multicol}
\usepackage{youngtab}
\usepackage{lipsum}
\usepackage{mwe}

\newcommand{\beq}{\begin{eqnarray}}
\newcommand{\eeq}{\end{eqnarray}}
\newcommand{\bea}{\begin{eqnarray}}
\newcommand{\eea}{\end{eqnarray}}
\newcommand{\be}{\begin{equation}}
\newcommand{\ee}{\end{equation}}

\def\brc{\langle}
\def\ckt{\rangle}

\def\de{\partial}

\def\Tr{\qopname\relax o{Tr}}

\numberwithin{equation}{section}

\numberwithin{equation}{section}

\begin{document}
 

\title{On the negative-result experiments \\  in quantum mechanics  }  


\author{  
 Kenichi Konishi$^{(1,2)}$   \\[13pt]
 {\em \footnotesize
$^{(1)}$INFN, Sezione di Pisa,    
Largo Pontecorvo, 3, Ed. C, 56127 Pisa, Italy}\\[2pt]
{\em \footnotesize
$^{(2)}$Department of Physics ``E. Fermi", University of Pisa,} \\[-5pt]
{\em \footnotesize
Largo Pontecorvo, 3, Ed. C, 56127 Pisa, Italy}   \\[2pt]
\\[1pt] 
{ \footnotesize 
kenichi.konishi@unipi.it,      
orcid.org/0000-0002-4944-5444 }  
}
\date{}


\maketitle

\begin{abstract}
{\footnotesize  We comment on the so-called negative-result experiments (also known as null measurements, interaction-free measurements, and so on) in quantum mechanics (QM),  in the light of  the new general understanding of the quantum-measurement processes, proposed recently.  All experiments of this kind (null-measurements) can be understood as improper measurements  with an intentionally biased detector set up, which introduces exclusion or selection of certain events. 
 The prediction on the state of a microscopic system under study  based on a null measurement,  is  sometimes dramatically described as  ``wave-function collapse without any microsystem-detector  interactions".   Though certainly correct, such a prediction is just a consequence of the standard QM laws,  not
  different from the situation in the  
 so-called state-preparation procedure.  Another closely related concept is the (first-class or) repeatable measurements.  
 The verification of the prediction made by a null-measurement requires eventually a standard unbiased measurement involving the microsystem-macroscopic detector interactions, which are nonadiabatic, irreversible processes of signal amplification.}

\end{abstract}

~~~{\footnotesize {\bf Keywords:} ~ {Particles, Quantum measurement, Null measurement, Wave-function collapse}}

\bigskip

\newpage

\tableofcontents

\newpage

\section{Introduction}

A typical quantum-mechanical measurement process involves an interaction between a microscopic quantum system with a macroscopic experimental device,  which is capable of  faithfully capturing the quantum state of the microscopic system - object of the ``measurement" -  and of recording  the result  in the form of a classical state of matter.  The process  typically involves  a non-adiabatic,  irreversible process of  signal amplification (such as a chain ionization and fixture of images on a photographic film).   Critical discussions on earlier attempts for constructing a theory of measurements can be found in \cite{Joos};  many original papers are collected  in \cite{WheelerZ}.  See also \cite{Bell,Peres}.
Quantum measurement processes have been analyzed recently, with a few new key observations \cite{KK,KKTalk}.

Such a physical characterization of a quantum measurement was challenged by a series of gedanken (as well as real) experiments   \cite{Renninger,BombTester}   of particular kind,  in which some negative-result (or null) measurement, hence without any microsystem-macroscopic-device interactions typical of the standard  quantum measurements, allows one to acquire a nontrivial  information on the quantum state of the system under study.  The predicted state,   $ |\psi^{\prime} \ckt$,  is necessarily a restriction \footnote{Namely,  a projection of the original vector in the Hilbert space onto a  vector in a space of a smaller dimension.}   of the original wave function $|\psi\ckt$.  This fact was somewhat dramaticaly expressed as  ``an interaction-free measurement   leading  to a wave-function collapse".   
These negative-result-experiment arguments have been presented, and believed by some,  as counter-examples to the physical characterization of  the quantum measurement processes,    i.e.,    
non-adiabatic,   irreversible processes involving  micro system - macroscopic device interactions.  

The aim of this short note is twofold.  The first is to review the characteristics of a typical quantum-measurement processes, based on a few novel observations made recently \cite{KK,KKTalk,KK2}.
The second  is to give a simple, more sober interpretation of these null measurements:
   they all correspond to specific, biased detector set-ups  which  select or exclude certain possible events.
    The prediction made on the quantum state after a null-measurement,  certainly correct,  is  just a consequence of the standard quantum mechanical laws: they do not in any way disprove the physical picture of a typical quantum measurement process.  Indeed, verification of the predictions made by a null-measurement on the system being studied, eventually requires a proper, unbiased  quantum measurement, involving microsystem-macroscopic-device interactions. 

Even though the main subject of this note \cite{Renninger,BombTester}  is rather an old one (see however  \cite{Aguilar,Vaidman2} for recent
discussions),  and has been discussed extensively (see \cite{Joos} and the references cited in \cite{Aguilar,Vaidman2}),  it touched some of the most subtle aspects of the interpretations of the quantum mechanical predictions.
It is thus not entirely pointless today, perhaps, to revisit the essential aspects of these negative-result experiments, correct any inappropriate interpretations, and 
 to ensure that our understanding of the QM laws be crystal clear.


\section{Solution of the quantum-measurement problem  in a  nutshell  \label{processes}   }

A measurement of a quantity $F$,    done on a quantum state $\psi$      
\be    |\psi \ckt =   \sum_n  c_n  | n \ckt \;, \qquad   \sum_n  |c_n|^2=1\;, \label{state}   
\ee
where  $|n\ckt$ is the eigenstate of $F$ with eigenvalue, $f_n$,    
 used to be schematized as
 \bea     |\psi \ckt \otimes   |\Phi_0 \ckt    \otimes   |Env_0  \ckt   &=&    \left( \sum_n  c_n  | n \ckt  \right)        \otimes   |\Phi_0 \ckt    \otimes   |Env_0  \ckt  
\label{formula00}    \\  & \longrightarrow &   \sum_n  c_n  | n \ckt        \otimes   |\Phi_n \ckt    \otimes   |Env_0  \ckt   
\label{formula11} \\  & \longrightarrow &       \sum_n  c_n  | n \ckt        \otimes   |\Phi_n \ckt    \otimes   |Env_n  \ckt      \;,   
\label{formula22} \eea 
where $ |\Phi_n \ckt$ represents the detector state, with has recorded the result  $F=f_n$ and  $ |Env  \ckt $  stands for 
everything else,  the state of the experimentalist and the rest of the world. The index $0$ indicates a neutral state, whereas the index $n$ stands for the measurement result, $F=f_n$. 
Such a formula is found in many textbooks of quantum mechanics (QM), and (basically) in all past discussions     \cite{Joos,WheelerZ}.

   Actually,   the formulae  (\ref{formula00})-(\ref{formula22}) are incorrect  in several accounts  \cite{KK,KKTalk}.   
   
     (i)     First, the factorized form for  the wave function expressed by  the symbol $\otimes$ in the state after the measurement  is not valid  \footnote{As for the state before the measurement in the first line,  (\ref{formula00}), factorization of $|\psi\ckt$   is  correct:  it must be so in any ideal experiment.  Factorized form $ |\Phi_0 \ckt    \otimes   |Env_0  \ckt  $  is instead incorrect, see below and \cite{KK}.    }.   In general, in the aftermath of a measurement  the microscopic system gets entangled with the device and with the environment, typically in an uncontrolled manner \footnote{In an exceptional  class of the so-called repeatable  (or the first-class) experiments,  the microscopic system under study remains 
  factorized and intact.  These processes are indeed closely related to the ``negative-result experiments''  as well as  with the state preparation procedure,
 as discussed below,   Sec.~\ref{negative}.
  }, even though the information about its original state is faithfully recorded by the detector.

(ii)     Secondly,  the experimental device,  $|\Phi\ckt$,  typically a macroscopic body at finite temperatures, and entangled with the rest of the world (environment),  is 
itself in a decohered, mixed state  \cite{Joos1}-\cite{Hansen},\cite{KK2}.     This is so even before the measurement,  
let alone  after the measurement. 
  This means that 
the  expression (\ref{formula11}) should not be considered as a pure state (a coherent superposition):   it is a mixture \footnote{This observation is sufficient to explain away  a ``puzzle''   recently discussed in \cite{Renner}. }.

Taking these points into account,   it was proposed  \cite{KK} that the detector-environment entangled state be denoted  as $ |\Phi;    Env \ckt$.
The typical measurement process thus looks more like
   \bea     |\psi \ckt \otimes |\Phi_0;  Env_0 \ckt  &=&   \left( \sum_n  c_n  | n \ckt  \right)    \otimes   |\Phi_0;    Env_0 \ckt\,,     \qquad {\rm (before)}  \label{step000}  \\
   & \longrightarrow &       \sum_n  c_n  \,    |{\tilde  \Phi}_n;     Env_0 \ckt \,,     \qquad {\rm (after)}        \label{step111} \\
  & \longrightarrow &     \sum_n  c_n   \,   |{\tilde  \Phi}_n;     Env_n \ckt\,,     \qquad {\rm (later)}       \;,\label{step222} 
 \eea
 where $ |{\tilde  \Phi}_n;     Env_0 \ckt$   with   the symbol ${\tilde  \Phi}_n$    stands for the microsystem-detector entangled  state  (see \cite{KK} for more discussion)  with a macroscopic marking of the recording,  $F= f_n$.

  The essential features of the process, (\ref{step000}) $\to$   (\ref{step111}), are the following  \footnote{The second stage of the process,  (\ref{step222}),  in which  the 
experimentalist sees the result of the measurement on her computer screen,  the others read about it in Physical Review, etc., is
totally irrelevant, in spite of many sophisticated and sometimes philosophical discussions made in the past (see \cite{WheelerZ}   for some).}.  
  
    \begin{description}
    
  \item[(a)] {\it  Each}  term in (\ref{step111}) containing  $  |{\tilde  \Phi}_n; Env_0 \ckt $ is a complicated mixed state   (point (ii) above), representing the microsystem-detector-environment entangled state  (point (i) above),   with a well-defined macroscopic marker  of the 
  measurement result, $F= f_n$.  It is an eigenstate  of the operator $F$. Namely, 
      \bea      
    F\, | n \ckt  &=&   f_n\, | n \ckt     \label{before}  \\
     \to \qquad  F\,    |{\tilde  \Phi}_n;     Env_0 \ckt  &=&   f_n    |{\tilde  \Phi}_n;     Env_0 \ckt \,.  \label{after}
      \eea
      The relation between (\ref{before}) and  (\ref{after})  {\it  defines} a good, faithful measurement.

  \item[(b)] A key observation  \cite{KK}  is  that, reflecting the pointlike nature of the fundamental entities of our world, 
    each measurement process is a spacetime pointlike event 
  (or triggered by one).     This entails that   {\it   the wave functions corresponding to the different terms in  (\ref{step111}) have no overlapping spacetime supports.}
  Thus,  not only the orthonormality
  \be              \brc    {\tilde  \Phi}_m;     Env_0                |{\tilde  \Phi}_n;     Env_0 \ckt   = \delta_{mn}  \,,\label{Orth} 
  \ee  
holds,     but also a dynamical diagonalization 
   \be              \brc    {\tilde  \Phi}_m;     Env_0      |     G      |{\tilde  \Phi}_n;     Env_0 \ckt   =   G_n   \,  \delta_{mn}  \,.\label{diagonal11}
  \ee  
occurs  for {\it  any} local operator $G$.   
  Note that  $G_n$  in  (\ref{diagonal11}) is defined by  (\ref{diagonal11})  itself: it is unrelated to the eigenvalues of the operator $G$
 in the isolated microscopic  system before the experiment.

 The diagonalization  (\ref{diagonal11})  is of utmost importance.

 {\it   Before  the measurement,}  in the state (\ref{step000}),  the expectation value of a generic quantity  $G$ is given by
 \be   \{ \brc \Phi_0;  Env_0 |  \otimes     \brc  \psi | \}   G    \{ |\psi \ckt \otimes |\Phi_0;  Env_0 \ckt \}=
   \brc  \psi |  G  | \psi \ckt=  \sum_{m,n} G_{mn} c_m^* c_n  \equiv \Tr  \rho^{(0)}   G\;,  
 \ee
 ($G_{mn}\equiv  \brc m | G | n \ckt$)    meaning that the system is described by a density matrix 
 \be    \rho^{(0)}_{nm}   =        c_n   c_m^*\;,    
 \ee
i.e., by  the pure state,   $|\psi\ckt=  \sum_n c_n |n\ckt.$   For $G=F$, the variable whose eigenstates are taken as the basis  $\{ |n\ckt \}$,   one finds of course the  standard formula
\be  \brc  \psi |  F  | \psi \ckt= \sum_n f_n |c_n|^2\;.  \label{standard}  \ee
 
{\it  After the measurement, }   according to  (\ref{diagonal11}) the expectation value  of a generic variable $G$   taken in the ``state"  (\ref{step111}), is given 
(by using (\ref{diagonal11})) by
 \be              \left( \sum_n  c_m^*  \,    \brc      {\tilde  \Phi}_m;     Env_0  |   \right)  \,     G \, \left( \sum_n  c_n  \,    |{\tilde  \Phi}_n;     Env_0 \ckt \right)= \sum_n |c_n|^2  G_n\;. 
 \label{ExPect}
\ee 
That this holds for {\it any}  operator $G$ means that the density matrix of the system  has been effectively reduced to a diagonal form
  \be  \rho^{(0)}    \stackrel {(measurement)} { \Longrightarrow }   
   \rho^{(1)}   =    \left(\begin{array}{ccccc}|c_1|^2  &     &    &  &  \\    & |c_2|^2 &  &  &  \\      &  & \ddots &  &  \\  &  &  & |c_n|^2 &  \\ &  &  &  & \ddots\end{array}\right)\;.\label{diagonal}
\ee 
By paraphrasing the ``environment-induced superselection rule"  \cite{Joos}, we may call (\ref{diagonal}) the measurement-induced superselection rule. 
  
    \item[(c)] 
 The fact that the wave functions of the different terms in (\ref{step111}) have  no overlapping spacetime support  means that    
the aftermath of each measurement event  is a single term in (\ref{step111}) \footnote{A related fact is that the detector-environment  ``state"  $ |\Phi_0;  Env_0 \ckt$, even if  it might look identical macroscopically, it can never be the same quantum state at two different measurement instants.  The time evolution of the macroscopic number of molecules and atoms in the detector and environment, means  that the   ``state" just before each experiment $ |\Phi_0;  Env_0 \ckt$ is a unique and distinct quantum state, actually carrying a hidden index ``(n)" of each measurement. },  
that is,  the time evolution in {\it   each single measurement}   is, 
\be    \left( \sum_n  c_n  | n \ckt  \right)    \otimes   |\Phi_0; Env \ckt    \Longrightarrow     |{\tilde \Phi}_m ; Env \ckt \;:      \label{JumpCor}
 \ee
 i.e.,  with a single term present  the instant after the measurement (e.g., with $F=f_m$).    
(\ref{JumpCor})  is often (improperly)   described as a   ``wave-function collapse".  \footnote{
The words evoke in our mind an image of some distribution suddenly contracting, which does not correspond to any real physical process. The wave function is itself not an observable. 
   }
   
    \item[(d)] A second crucial consequence of our  description of the measurement process,  (\ref{step000}), (\ref{step111}), concerns    the {\it  repeated}  
 measurements.  For the measurement of the quantity $F$,   it follows from (\ref{after}), (\ref{Orth})  and (\ref{ExPect})  that   the expectation value is given by 
 \be              \left( \sum_n  c_m^*  \,    \brc      {\tilde  \Phi}_m;     Env_0  |   \right)  \,    F   \, \left( \sum_n  c_n  \,    |{\tilde  \Phi}_n;     Env_0 \ckt \right)= \sum_n |c_n|^2  f_n\;, 
 \label{ExPectF}
\ee 
where $f_n$  {\it  are}   the eigenvalues of  $F$.  This  means that
the relative frequency for finding the result, $F= f_n$,   has been found to be given by 
\be 
{\cal P}_n = |c_n|^2\;.   \label{rule}  
\ee
\end{description}

  {\it  \noindent   The derivation of  the  ``wave-function collapse"  (\ref{JumpCor})  and of the formula for the relative frequency  (\ref{rule}), amount to  the solution of  the quantum-measurement problem. }
  
 One might object that we have  just reproduced the standard Born rule. This is not  quite so.  
Unlike the latter, our description explains why and how the ``wave-function collapse" occurs, and yields the rule (\ref{rule})  
 as the result of physical measurement processes involving the microsystem-detector-environment interactions. 
Even though they look similar, the difference in the meaning of (\ref{standard}) and (\ref{ExPectF})  is crucial.
  Last,  by  eliminating the fundamentally obscure concept of ``probability" inherent to Born's rule,  and by replacing it by the (normalized) relative frequency for various outcomes in repeated experiments, it leads to a more natural  interpretation of  the QM laws \cite{KK,KKTalk}.   For instance, the concept of the ``wave function of the universe" makes perfect sense now,  whereas  in the traditional interpretation of QM based on Born's rule one cannot  avoid falling into a conundrum of having nobody outside the universe, observing it and making repeated experiments on it.    Note that   the cosmologists today 
are adopting this new interpretation of QM laws, naturally,  when they discuss the structure formation in the early universe, through the density quantum fluctuations at certain stage of the inflation.

%

 \subsection{A secret key} 
 An alert reader must have noticed the following  subtlety.   We affirmed that each of the microsystem-detector-environment  entangled  state $  |{\tilde  \Phi}_n;  Env_0 \ckt $  is a complicated mixed state,  involving the myriad of microscopic processes, such as the scattering of air molecules against the   $\sim O(10^{25})$ atoms and molecules composing the detector, the emission of the infrared photons from the latter,  and so on. Nevertheless, we used it as an ordinary wave function (i.e.,  a pure state),  to evaluate the expectation values, (\ref{ExPect}),  (\ref{ExPectF}).  Is it consistent?        

Actually here lies a secret key in the whole discussion. What might not be widely appreciated is the fact  that there are no differences of  principle between the concepts of the pure and mixed states. Consider any  (pure) quantum state
 $\Psi( \{{\bf \xi}_i\},  \,\{ {\bf \chi}_k\})$  where  ${\bf \xi}_i\equiv  ({\bf r}_i, s_i)$  and   ${\bf \chi}_k   \equiv ({\bf r}^{\prime}_k, s^{\prime}_k)$  are the position and spin component of the particles composing the whole system.   We (the physicists) are assumed to have access only to the subsystem  ($A$)   containing the degrees of freedom  $ \{{\bf \xi}_i\}$.  The rest of the world ($B$) described by  $ \{{\bf \chi}_k\}$ is off-limit.   By introducing an orthonormal (ON)  set of states describing the system $A$, 
 $\{ |n\ckt^{(A)}  \}$, and  expanding the coefficient  in ON  states of $B$,   $\{ | N  \ckt ^{(B)} \}$,  a generic state has the form, 
 \be    |\Psi\ckt= \sum_{n, N}  c_{n,  N}  \, |n\ckt^{(A)}  |N\ckt^{(B)} \;.  \label{WF}
 \ee
 The expectation value of any variable $G$ pertinent to $A$, is then
 \be   \brc \Psi | G |\Psi \ckt       = \Tr   {\bf G} {\rho} \;,  \qquad    {\bf G}_{mn}=   \brc m | G | n \ckt\;, \quad     \rho_{nm}    =    \sum_N   c_{n,N}  c_{m, N}^*   
 \ee  
 where $\rho$ is the density matrix.  It is perfectly correct however to use the wave function (\ref{WF}), or (\ref{step111}),  e.g., in (\ref{ExPect}) or  in (\ref{ExPectF}),  when the sum over the system $B$ is indeed  implied in the calculation \footnote{A  similar idea was used in \cite{KK2} to explain why the Ehrenfest theorem can be used to derive Newton's equations for the center of mass of a macroscopic body at finite body temperatures, which is in a decohered, mixed state.  }.  
 
To sum up, the concept of a mixed state (versus a pure state)  is a relative one, depending on which part of the world  ($A$)  is accessible to us.
 The ignorance about the rest of the universe ($B$)   is parametrized by the density matrix.  More interestingly,  two (or more) groups of physicists  studying  a common event such as a multiparticle decay,  located in spacetime regions which are spacelikely separated,  might find  apparently paradoxical outcomes involving various quantum correlations.    These phenomena of   ``quantum nonlocality", however  as fascinating and mysterious they might continue to appeal to human mind, 
 are today fully understood.    The peculiar  ``subjectivity"  of the quantum-mechanical laws  also hinges upon these circumstances.

%
 
%
%

\subsection{A remark}   
    One of the characteristics (or requirements) of a proper quantum measurement, which every experimentalists know well,   is that  the device must not have any  bias,  i.e., it should be equally effective to register all possible outcomes, $f_n$.  For, if it were not so, the experimental average for the frequency times various possible results would not match the theoretical prediction, (\ref{rule}), even after many repeated measurements.   

An exception occurs  when there are only two  possible outcomes, either $F=f_1$ or $F=f_2$.  In this case, the device which is capable of measuring only one of the 
possible results, e.g., $f_1$ (the so-called yes-no experiment) is sufficient  to give unbiased measurement results.  Event by event,  the detection of  $f_1$ 
(yes) means the  wave function collapse (\ref{JumpCor}) with $F=f_1$;   the non detection of  $f_1$  (no)  implies that the state is  $|\psi\ckt= |2\ckt$,   even if the measurement  $F=f_2$  has not yet been actually  done.   The negative-result experiments to be reviewed below rely on an analogous logic.

\section{Negative-result experiments  \label{negative}} 

In this section a few well-known examples of the negative-result experiments are reviewed and their essential features critically analyzed.

\subsection{Renninger    \label{SecRenninger} }  

A pair of collections of particle detectors,  each covering one of the hemispheres, surround  a radioactive nucleus \footnote{In the original article by Renninger \cite{Renninger},  an excited atom  
is used and what is emitted is a photon. Nothing essential changes by replacing the atom in an excited level by an unstable nucleus, however.}  in the center,  which emits an $\alpha$ particle, e.g., in S-wave.
If one of the shells does not observe $\alpha$,   its wave function has been ``collapsed to the other hemisphere",  without any interactions between the $\alpha$ particle and the experimental device!

To make the puzzle look sharper, the detectors  in the second hemisphere   (call $\Phi^{(lower)}$) may be set at a radius much larger than the first. 
If  the half-life-time  of the nucleus  is $\tau$,  the nucleus has most likely decayed ($\alpha$ has been emitted)  
by the time  $t=  30  \tau$ \footnote{Only in an exceptional one out of    $e^{30} $ repeated experiments, on the average, the nucleus will be found still undecayed, 
without $\alpha$ emission. }. 
If the second detector is set at the distance 
\be   R >    30  c  \tau  \;,      \label{far} \ee
then by the  time $t= 30 \tau$, if the first detector has not detected $\alpha$, then  most likely it is still travelling towards the lower-hemisphere. 
So  it might appear as if ``the wave function had collapsed", without any particle-detector interactions having taken place.  
The ``paradox" is only apparent, and its origin can be traced to the misconception that the wave-function collapse is a sort of real  physical process in itself, as  noted
already.
%

The metastable nucleus  which  $\alpha$-decays   in an $S$ state,   can be expressed by
\be       |\Psi \ckt =      |\Psi^{(0)} (t)  \ckt   +      |\Psi^{\prime} (t)  \ckt   |\alpha \ckt  \;,  \label{above} 
\ee
At the time  $t=  30  \tau$  the nucleus has most certainly decayed, so let us concentrate on the second term of (\ref{above}). 
 The $\alpha$ particle is described by an S-wave function \footnote{This is really the reduced one-body description of the $\alpha$ particle -
nucleus. 
  }, 
\be         |\Psi \ckt  =     \sum_i   c_i   |  i \ckt       \;, 
\ee  
where $i=1,2,\ldots  N $, $N \gg 1$,  represent the uniformly  discretized cells of the  $4\pi$ solid angle. 
 The S-wave nature of the wave function means that
\be       \forall i \qquad      c_i=    \frac{1}{\sqrt{N}}\;.   
\ee         
This is simply a discretized version of the $S$-wave  wave function,  
\be  \Psi({\bf r}) =  f(r)\sim  \frac{e^{ikr}}{r}\;,    \label{Swave} 
\ee
independent of the angular variables. 
The standard  measurement of the angular distribution is done with 
\be         |\psi \ckt \otimes  |\Phi_0\ckt   = \sum_i   c_i   | i \ckt  \otimes  |\Phi_0  \ckt   \;.    \label{thestate} 
\ee  
where the detector   $\Phi_0$  is uniformly sensitive  over the $4\pi$  solid angle.  It should not have any bias as to which angular direction the particle is eventually 
measured.


The experimental set-up  by Renninger,  instead,  has  a nonuniform  $\Phi_0$.    Namely,  
\be       |\psi \ckt  |\Phi_0\ckt   = \sum_{i=1}^{N/2}   c_i   | i \ckt  \otimes  |\Phi_0^{(upper)}  \ckt    +    \sum_{i=N/2+1}^{N}   c_i   | i \ckt  \otimes  |\Phi_0^{(lower)}  \ckt     \;.  
\label{upperlower}  
\ee
where  $\Phi_0^{(upper)} $ and $\Phi_0^{(lower)}$   are the first and second  groups of detectors at the upper and lower hemispheres, respectively.
The detector   $\Phi_0^{(upper)}$ ($\Phi_0^{(lower)}$)  is insensitive to the $\alpha$ particle flying towards the lower (upper) hemisphere,  thus 
the cross terms such as  $ \sum_{i=1}^{N/2}   c_i   | i \ckt  \otimes  |\Phi_0^{(lower)}  \ckt  $  are absent.

 In the case  (\ref{far}) the two detectors are such that   up to $t=  30 \tau$, the lower detector is insensitive to the $\alpha$ particle, i.e., is not able to detect it. 
 Only the first is sensitive, and only to the $\alpha$ particle  travelling towards the upper hemisphere.

If the experiment is repeated,    half of the times    the detector  $\Phi_0^{(upper)} $  will record  $\alpha$  at time $t\le  30 \tau$,  and the other half of the times 
 it will not.   This is  what this particularly  biased detector will produce, in accordance with the QM prediction.   
 
 In a second type of events, where the first detector   $\Phi_0^{(upper)}$  does not detect  the $\alpha$ particle
(``the negative-result measurement"), 
  it is correct  {\it   for us to infer  that  }  the wave  function is reduced to the scond term of (\ref{upperlower}),    
  \be       |\psi \ckt  |\Phi_0\ckt   \to     \sum_{i=N/2+1}^{N}   c_i   | i \ckt   \otimes |\Phi_0^{(lower)}  \ckt     \;.  
\label{loweronly}  
\ee
In these events the $\alpha$ particle will be detected by  $\Phi_0^{(lower)}$   at a later time,  
 \be     t \sim   \frac{R}{c}  \gg   30 \tau\;, 
 \ee
 as will be  eventually verified by the standard  $\alpha$ - $\Phi_0^{(lower)}$ detector   interactions. 
 
The negative-result experiment such as this could look somewhat paradoxical.
   A phrase such as  ``for us to infer that ..." might  indeed appear to indicate  that  the ``wave function  collapse," $ |\psi \ckt   \to     \sum_{i=N/2+1}^{N}  | i \ckt\,,$  has been caused by human mind - the realization of the non-detection fact.    
   The wave-function collapse, so it might seem,  does not need any microsystem-macroscopic device interactions as those assumed in \cite{Daneri,Rosen}.      
     These questions were at the center of ardent debates, partially reignited by Renninger's work   (see for example,    \cite{JWY}, \cite{Loinger}).    

   What happens actually  in the Renninger experiment  is that,  in each decay event,  the $\alpha$ particle is  emitted  either towards the upper hemisphere or towards the lower hemisphere.
    With the same relative frequencies, if the experiment is repeated.  That is all.

   The aim of revisiting these old, and after all simple,  issues nonetheless,  was to illustrate how the wrong wordings and the misconception about the ``wave-function-collapse"
    have led to nonexistent, and hence unsolvable, problems in the past discussions on QM.

\subsubsection{State preparation  \label{statepr} } 
 The deduction (\ref{loweronly}) in the case of Renninger's negative-result experiment, is not essentially different from the preparation of a collimated atomic beam,  by using two successive slits,  such that 
the particles which have passed  both slits have a more or less well-defined momentum direction, so that one can  predict that the particle, left freely propagating,  will be detected in the direction of a straight line connecting 
 the two successive slits,  within some error  (taking into account the diffraction effects).  

The analogy may be made even closer,  by making the  upper detector $\Phi^{(upper)}$ of Renninger  cover 99 \% of the  $4\pi$  solid angle, leaving a small hole in   the south-pole direction.  Then in the  very rare event   (one in $100$)    in which 
   $\Phi^{(upper)}$ has not recorded  $\alpha$,    it can be predicted that the $\alpha$  will be  detected in the direction of the south-pole  direction 
   by $\Phi^{(lower)}$,   later.  
     Instead of the beam preparation by using two slits, here one uses just one slit and the selection of the (non-observation, null-measurement) event. 

\subsubsection{{$\alpha$} particle tracks in a cloud chamber  \label{Mott} } 

Renninger's process brings us back to one of the oldest ``puzzles"  in QM:  why  does an $\alpha$ particle 
emitted by a metastable nucleus, described by a spherically symmetric wave function, (\ref{Swave}),  produce   {\it  each}   
an (almost) straight-line track,  instead of a sequence of ionization blots distributed all over $4\pi$ angular directions?      
The answer has been given by a standard perturbation analysis made by Mott \cite{Mott},  and we are not here to discuss it anew. 

The reason why we brought up this historical issue here,  in spite of  little direct logical connection with the negative-experiment problems,   
is this.  The observation of an $\alpha$ particle track  in a Wilson chamber  is a quantum measurement of the angular distribution of its momentum.  
It is a {\it   measurement of intermediate type}, between the general one ((\ref{step111}) and (\ref{JumpCor})),  and  the repeatable experiments  ((\ref{repeatable00}) and (\ref{repeatable11}) below).   
In the latter, special type of experiments, the state of the microsystem under study remains factorized and intact as a pure state:  the measurement
process can be used as a state preparation.  

 In the former, more general type of measurements,  the microscopic system gets entangled with the device and with the environment in the process, and the information about its quantum state gets lost completely, in general, after the measurement event  (see (\ref{JumpCor})).
 
An $\alpha$ particle track in a cloud chamber starts when $\alpha$  hits the first atom, ionizing it.  The  $\alpha$- electron scattering process is a spacetime pointlike event, 
triggering the measurement event.
  The state of the $\alpha$ particle is only slightly affected by the $\alpha$-electron scattering.
The large mass ratio,  $m_{\alpha}/m_e   \sim 8000$,   means that the momentum of the  $\alpha$ particle  is almost unaffected  by the $\alpha$-electron scattering,  as is obvious from kinematics  and as explicitly verified by a concrete QM calculations \cite{Mott}, as it proceeds along an almost straight path, hitting and ionizing a sequence of atoms on its way.    

In conclusion, the explanation of the (apparent) wave function collapse, (\ref{JumpCor}),  i.e., that a measurement process is effectively a spacetime local event,   is valid also  in the intermediate type of measurements,  such as the Mott process,  momentum measurements by using the magnetic fields, particle tracks in the vertex detectors,  and so on.

\subsection{Elitzur-Vaidman  bomb tester}

\begin{figure}
\begin{center}
\includegraphics[width=6in]{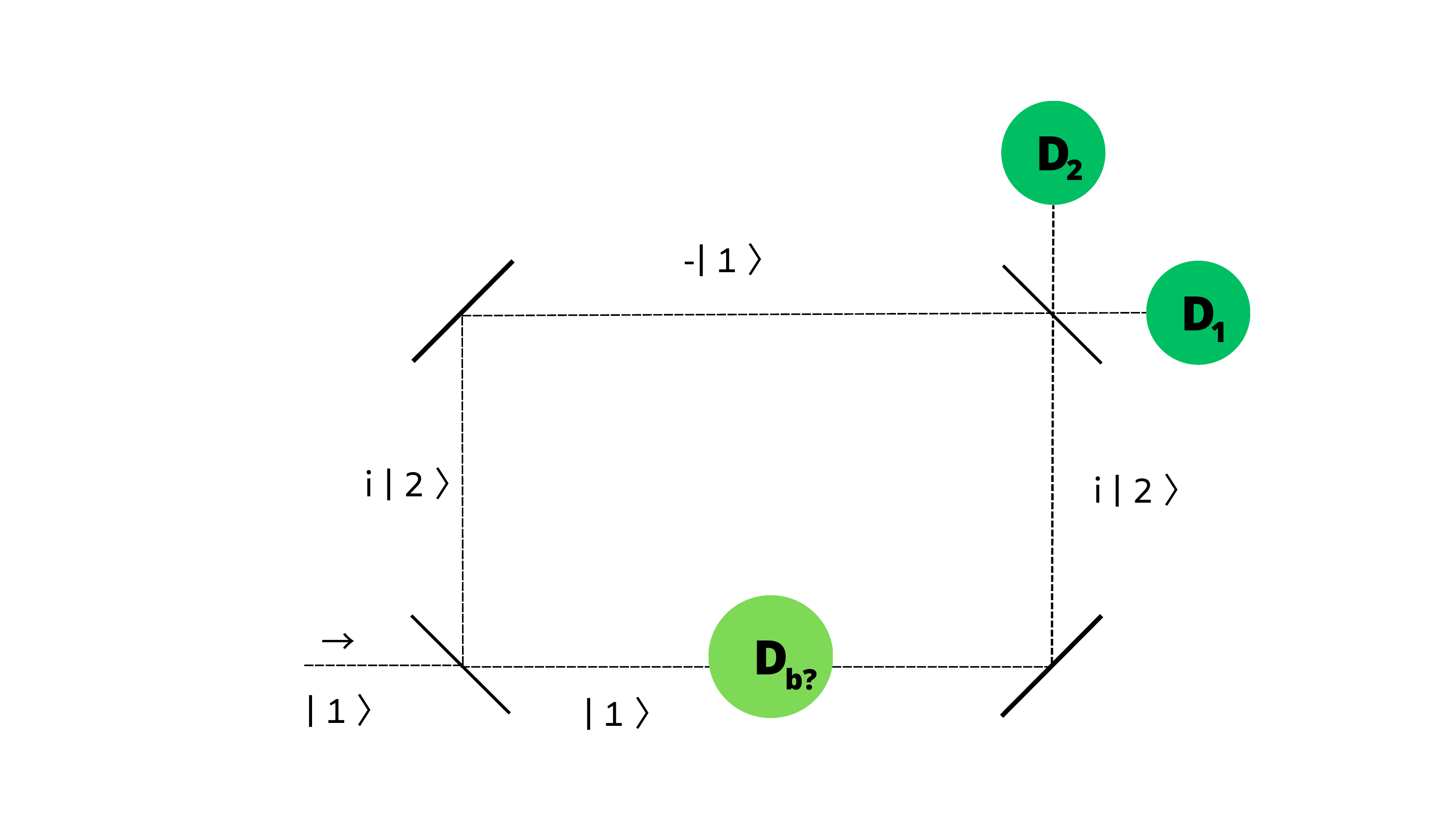}
\caption{\footnotesize  Elitsur-Vaidman bomb-tester experiment. The photon enters from the lower left corner to a Mach-Zehnder interferometry. The detection of the photon  at the detector  $D_2$ implies that the bomb is real, but that the photon has not interacted with the bomb.}
\label{Bomb}
\end{center}
\end{figure}

A more sophisticated,  amusing  set-up  is the so-called  Elitzur-Vaidman  bomb tester experiment \cite{BombTester}.
A single photon is sent to Mach-Zehnder interferometry.   See Fig.~\ref{Bomb}. 

A bomb which is either real or fake, is introduced in the lower path. 
After the passage through the first half mirror (the lower left corner in  Fig.~\ref{Bomb})   the original right moving photon $|1\ckt $  is converted to   
   the superposition 
\be  |1\ckt  \to    |\gamma\ckt=   \frac{1}{\sqrt 2} ( |1 \ckt   +  i  |2 \ckt ) \;,  \label{photon}
\ee
where $|2 \ckt$  is the wave packet of the photon  (reflected and)  moving upwards \footnote{We follow the notation and convention of \cite{BombTester}.  We recall that the photon, upon reflection,  acquires a phase shift of $\tfrac{\pi}{2}$.   }.

In the case the bomb is a fake \footnote{It is assumed \cite{BombTester}  that in that case the photon passes the region unaffected. }, the photon (\ref{photon}) 
goes through the two fully-silvered mirrors  (at the upper-left and lower-right corners of  Fig.~\ref{Bomb}), and is  in the linear superposition,
\be       |\gamma^{\prime}\ckt=   \frac{1}{\sqrt 2} ( i  |2\ckt   -   |1 \ckt ) \;,  \label{photon2}
\ee 
before entering the second half-silvered, beam splitter (in the upper right corner in  Fig.~\ref{Bomb}).   Going through it, the wave packet $|1\ckt$  gets transformed as in   (\ref{photon}), whereas  $|2\ckt$
goes into the state,
\be       |2  \ckt  \to    |\gamma^{\prime\prime} \ckt=   \frac{1}{\sqrt 2} ( i   |1 \ckt   +  |2 \ckt ) \;. \label{photonBis}
\ee  
Substituting (\ref{photon}) and (\ref{photonBis}) into   (\ref{photon2}),  we see that the state of the photon after the final beam splitter   is 
\be        |\gamma^{\prime\prime\prime}\ckt=  - |1\ckt\;,  \label{final}
\ee
which is purely right-moving.  Due to the interference effects the $ |2 \ckt $  component coming from the two terms of the original split-photon  state
(\ref{photon}), has canceled  \footnote{This is one of the beautiful features of the Eitzur-Vaidman experiment.  The interference effect,  characteristic 
of wave aspect of quantum particles, is here seen in a single photon event.  Typically   
the interference effects in QM,  instead, manifest itself after many identical experiments are repeated, such as in  \cite{Tonomura}.    }.  
Only the detector $D_1$  is triggered  (the photon detected)  by  $  |\gamma^{\prime\prime\prime}\ckt$, accordingly.

In the case it is a real one,   the bomb is a detector  inserted in the lower, horizontal section. 
 It is a quantum measurement  device to measure  the state of the photon in  $ |\gamma\ckt$,  
(\ref{photon}).   It is  however  a biased detector,    capable of registering only the photon travelling in the lower path,  $|1\ckt$  \footnote{It is thus  completely analogous to  the upper-half detectors in the  Renninger set-up, (\ref{upperlower}),  in which the second, lower-hemisphere detectors are set at a  large distance.  }.
There are two possible outcomes for each incident photon:    either  detection (explosion),  or non detection  (no explosion).    In the first case, 
the photon simply gets lost, and neither the detector $D_1$ nor   $D_2$  will register the photon. 

 In the second case  -  a sort of null measurement   -   
the wave function (\ref{photon})   is  reduced as  
  \be      |\gamma\ckt   \to  | 2 \ckt  \;,\label{No}   
\ee
in each such event.  The  photon gets then reflected by the mirror  at the upper-left corner  ($| 2 \ckt   \to i |1\ckt$), and  arriving at the final beam splitter,  gets  transformed  again as in (\ref{photon}).  
It is detected by detector $D_1$  half of the times, and by $D_2$   the other half of times, if the experiment is repeated.

In conclusion,   detection of the photon by detector $D_2$ implies that the bomb is real. The interesting point is that we know that this is so, but we know also that it has not exploded, i.e., the photon has not interacted with the bomb.   

Even though the phenomenon might look quite remarkable,  and is certainly not expected in classical physics,  everything follows from the standard QM laws.
If any,  as emphasized by Elitzur and Vaidman  themselves \cite{BombTester},  this process is interesting as a particular, peculiar  manifestation of quantum nonlocality.
The situation here might look rather different from the more familiar examples of quantum nonlocality associated with entangled  pairs of photons, electrons, etc.  
Actually, quantum nonlocality manifests itself whenever a microscopic system is in a pure quantum state with wave function having spatial support of a macroscopic extension   ((\ref{photon})  here).   As noted in \cite{KK},   quantum nonlocality is due to the absence in QM of any fundamental constant with the dimension of a length.

\subsection{Modified Stern-Gerlach set-up}  

The process  (\ref{No}) -  the negative-result event  -   can also  be regarded as  a particular realization of the so-called repeatable  experiment.
   A repeatable measurement is an exceptional class  of experiments in which the  
microscopic system  under study remains factorized (and intact) after the measurement, 
i.e., as 
\be     \left( \sum_n  c_n  | n \ckt  \right)        \otimes   |\Phi_0 \ckt 
  \longrightarrow    \sum_n  c_n  | n \ckt        \otimes   |\Phi_n \ckt   \;,
\label{repeatable00}
 \ee
 or focusing on a single experiment with the result, $F=f_m$,    
\be       \left( \sum_n  c_n  | n \ckt  \right)     \to  |m \ckt\;.   \label{repeatable11} 
\ee

A simple example of the repeatable measurement  is a variation of the Stern-Gerlach  (SG) experiment.   In the standard SG set-up (Fig.~2),  
an incoming beam of spin $\frac{1}{2}$ atom  (e.g., $A_g$),    travelling in the $\hat x$  direction, is  sent into a region of  inhomogeneous magnetic field, with a gradient,  
\be  \de B_z(z) / d z \ne 0\,.   \label{Bz}\ee   
  The incident wave packet, is divided into two,
\be   |\psi\ckt=   c_1 |\!\uparrow \ckt +  c_2 |\!\downarrow \ckt \;, \qquad  |c_1|^2+ |c_2|^2 =1\;, 
\ee 
with the spin-up wave packet  $ |\!\!\uparrow \ckt $  is deflected  upwards and the spin down component  $|\!\!\downarrow \ckt $  downwards, as the atom proceeds 
towards $\hat x$ direction.  On the screen they leave  
the two groups of blots  whose intensities (the numbers of atoms) are  proportional to   $|c_1|^2 :  |c_2|^2$ after many atoms have been registered.  

Though the Stern-Gerlach process is discussed in every textbook on quantum mechanics, there is some subtlety which is sometimes overlooked, due to the fact that the magnetic field satisfy Maxwell's equations  $\nabla \cdot {\bf B}=0,\,\, \nabla \times {\bf B}=0$.  The (apparent) puzzle is why, in spite of the fact that the condition  $\nabla \cdot {\bf B}=0$ implies that the inhomogenuity (\ref{Bz}) means an inhomogenuity ${\de B_y}/{\de y}\,$ of the same magnitude (if  $B_x=0$),  the net  effect is the deflection of the atom towards the $\pm {\hat z}$ direction only.  
The explanation  (the rapid spin precession and the cancellation of the forces in the $x-y$  directions)   with the discussion on the characteristics of the appropriate magnetic fields,  has been given in \cite{Platt,Alstrom}.

\begin{figure}
    \centering
    \begin{minipage}{0.45\textwidth}
        \centering
        \includegraphics[width=1.2\textwidth]{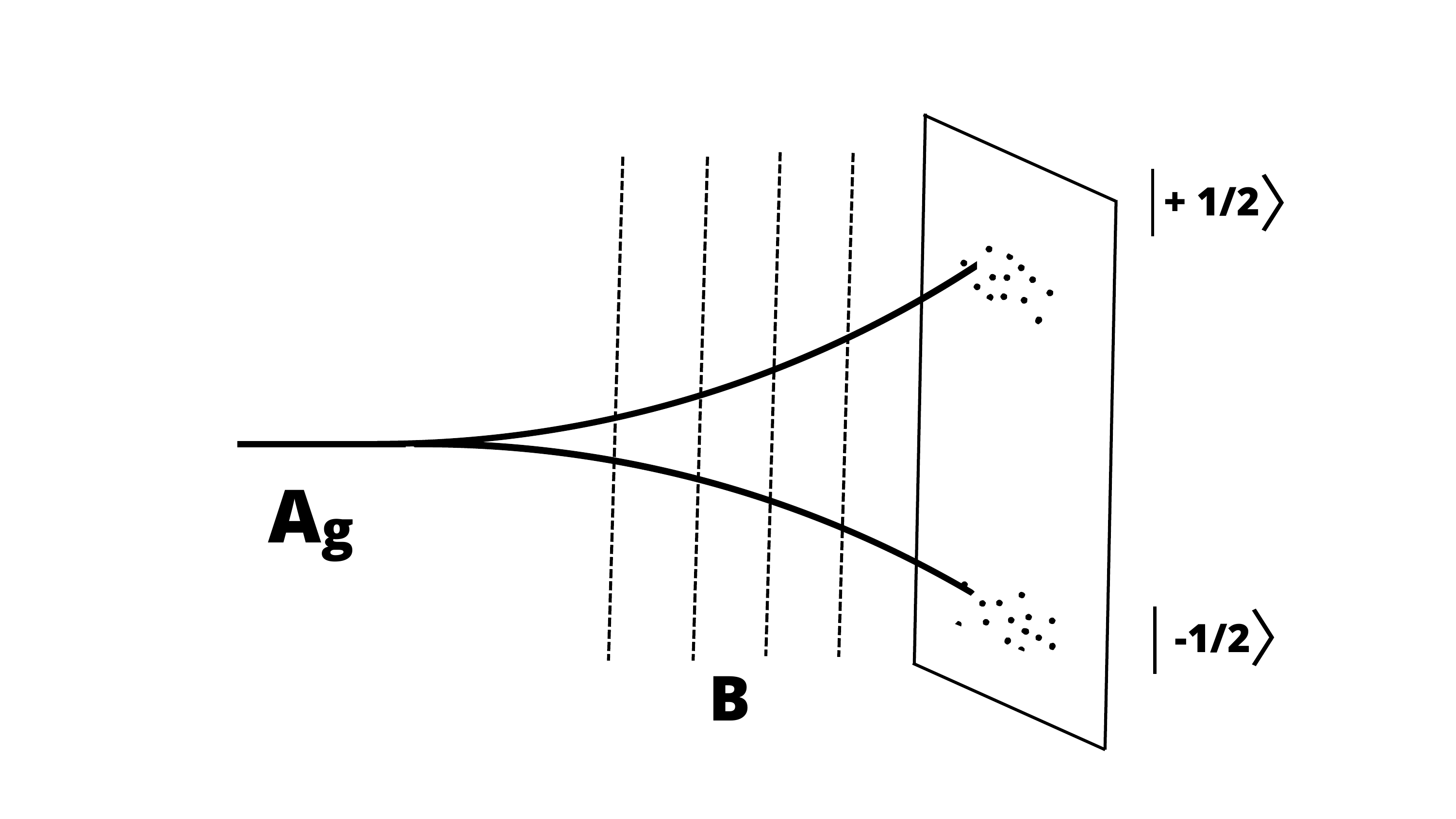} 
        \label{SG}
        \caption{\footnotesize  The standard SG set-up}
    \end{minipage}\hfill
    \begin{minipage}{0.45\textwidth}
        \centering
        \includegraphics[width=1.3\textwidth]{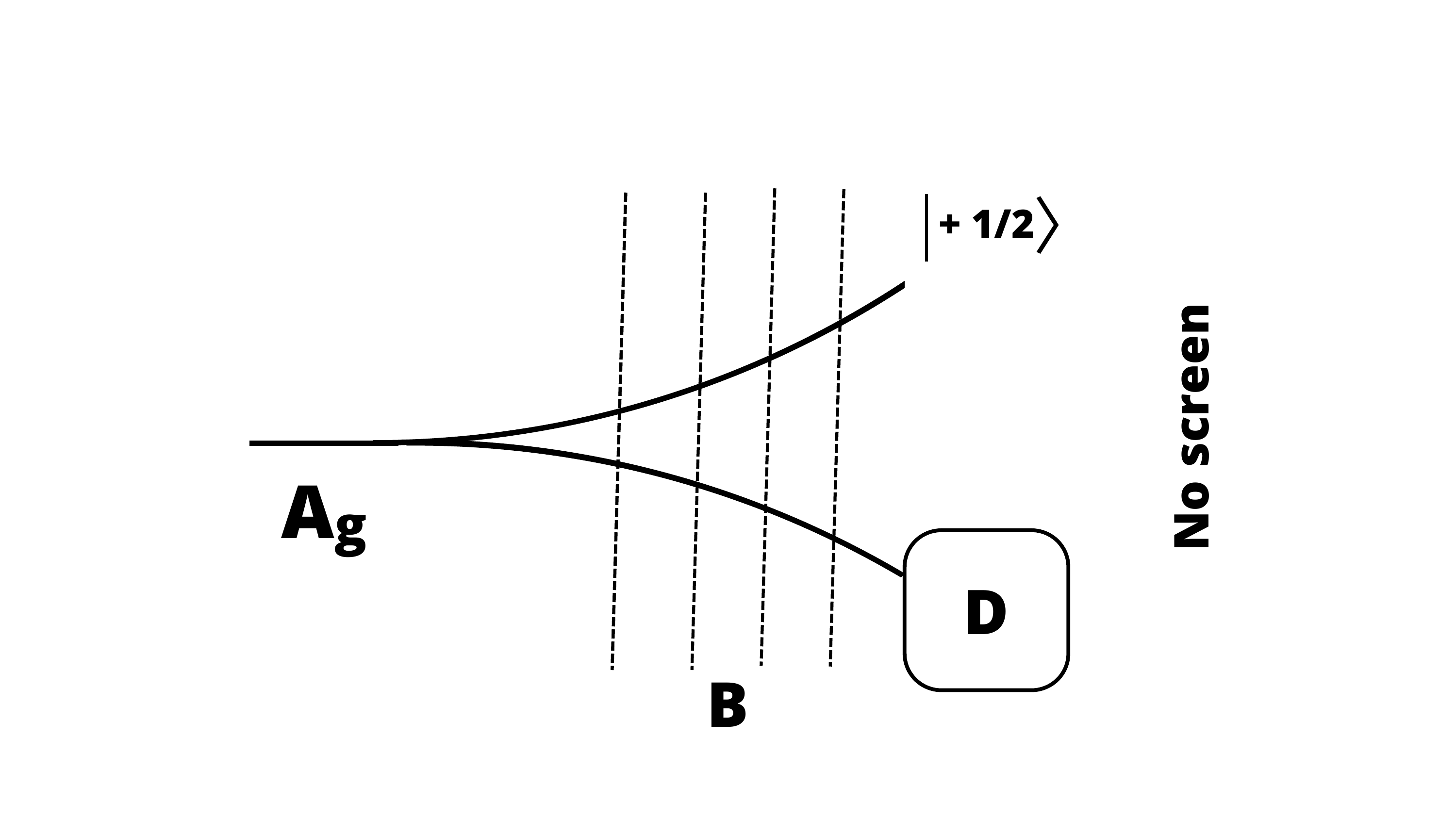} 
        \caption{\footnotesize The modified SG set-up}
         \label{SGvariation}
    \end{minipage}
\end{figure}

In a possible variation of the Stern-Gerlach set-up (Fig.~\ref{SGvariation}),   a  detector  $D$    is inserted in the region where the lower wave packet passes \cite{Loinger}.
The screen behind the region of the inhomogeneous magnetic field is eliminated.   
$D$  is analogous to the set of detectors in the upper-hemisphere in Renninger's set-up: it is a biased detector,  capable of 
capturing and recording only the atoms in the spin-down state.  
  For each single incident atom,  $D$ either registers it (yes) or does not (no).    
In the first case, the spin has been measured to be in the state $s_z= - \tfrac{1}{2}$,  but  the atom itself gets lost in the complicated atom-detector interactions. 

In the negative answer case (null measurement),  no atom-detector interactions have taken place;   nonetheless its spin state is determined to be  $s_z= + \tfrac{1}{2}$.  
The atom is in the pure $|\!\!\uparrow \ckt $ state, and it can be used as the initial condition for subsequent analyses, for instance, with another SG  set-ups with the magnets oriented in another direction, etc.

The whole discussion can be readily generalized to the case of atoms with spin $1$,  $\tfrac{3}{2}$, etc.   by appropriately enlarging the set of detectors, so as to extract  and  prepare the state of any chosen spin state $|s_z=m\ckt$, through interaction-free, null measurements.

  The  modified SG set-up  we considered in this section  can  thus  be seen  as a simple, prototype version of Renninger's negative-result  experiment  \cite{Loinger},    as an example of the repeatable measurement,  or   as a typical  ``state preparation" process,   
 illustrating well  the fact that  these three concepts are closely related to each other \footnote{We have already seen a similar connection also  in Sec.~\ref{statepr}. }.

\section{Reflections}

 It is essential,  in all negative-result experiments discussed in Sec.~\ref{negative},  that  a very weak flux of the incident particles is used,  such that processes with 
a   {\it  single  incident}  $\alpha$ particle, a photon, or an atom, are studied.  Also, the experimental control must be good enough  so that  the 
expected time of arrival of each particle at the (biased) detector, is known with a reasonable precision.   
 The reason is that,  if it were not so,  the non observation of certain event would not lead to any useful conclusion,  as e.g.,  the particle may not yet have arrived,  or has already passed,  or is unknown when it will arrive,  and so on.   
 
{\it    In other words,  an ideal  null-measurement is a spacetime local event,  albeit a virtual (i.e., missed) one \footnote{In the Renninger experiment,  
even though the spontaneous $\alpha$ emission {\it  is} a spacetime local event \cite{KK},  the exact  instant  $\alpha$ is emitted cannot be predicted,  being a manifestation of quantum fluctuations.   This is the reason why one must construct the argument  by considering a lap of time (e.g.,   $t \le    30   \, \tau$), to make sure that the nucleus has decayed and  the $\alpha$ emission has taken place -   with certainty,  $1 - O(e^{-30})$.  
}}.
It  represents the other side of the same medal of 
the standard quantum measurement processes each of which is   
  a local spacetime event at its core  \cite{KK}.   
  This latter fact is the origin of the apparent ``wave-function collapse", as reviewed in  Sec.~\ref{processes}.

Another important reflection is that the discussions of Sec.~\ref{negative}  illustrate nicely the well known fact about QM, i.e.,   that the wave-like behavior  (the superposition, quantum nonlocality  and interference)  is the property of each single particle   (the $\alpha$ particle, the photon and the atom  discussed here,  or the electron in Tonomura's double-slit experiment \cite{Tonomura}), and not due to a collective motion of, or correlation among, the particles in the beam.   
 
The catchphrase  ``wave-particle duality"   was used historically to describe the apparently schizophrenic behavior of the electrons, photons  and atoms.    In hindsight, though,  this familiar expression left space for ambiguity and misunderstanding.  
For instance,   it is an entirely different story 
 that a large number of identical bosons form collective wavelike  motions,  such as the classical electromagnetic waves  (light),  or Bose-Einstein  condensed cold atoms which are described in terms of a macroscopic wave function.  So are all macroscopic quantum phenomena such as superconductivity
 and superfluidity, quantum Hall effects, and so on. 
   The wave-particle duality, a core idea of QM,  is the property of each single quantum particle \footnote{Indeed a less poetic  but more precise 
 expression would be  ``quantum fluctuations of a {\it  particle} described by the wave function."   The words ``particle" and ``wave" do not have the symmetric logical roles. }.

  
  A last consideration:  in this work we took it for granted that the $\alpha$ particle, the photon, and the silver atom, are all quantum particles. But what if a large molecule such as $C_{70}$  is used instead? Is it still a quantum particle?  See  \cite{C70,Arndt1} for some  developments in our understanding of these questions.  
  
The quest to grasp the very essential factors which can tell  quantum-mechanical particles  (the elementary particles, atomic nuclei,  atoms, small molecules, etc.)
 from classical ones  (the center of mass of isolated macroscopic bodies 
at finite body temperatures),  has led us recently  to the concept of the Quantum Ratio  \cite{KK2, KKHTE}. 

\section*{Acknowledgments} 
The author is deeply indebted to  Hans Thomas Elze,  Marco Matone, Pietro Menotti and Nobuaki Miyabe,  for information and for discussions.   This work is done under the support of the
INFN special project grant GAST (Gauge and String Theories).

\end{document}